\documentclass[pra,reprint,twocolumn,amsmath,amssymb,amsfonts,superscriptaddress]{revtex4}

\usepackage{graphicx}
\usepackage{dcolumn}
\usepackage{bm}
\usepackage{color}
\usepackage{amsmath,amsthm,amssymb,bm}
\usepackage{mathrsfs}
\usepackage{comment}

\newcommand{\R}{\mathbf{r}}
\newcommand{\K}{\mathbf{k}}

\newcommand{\p}{\mathbf{p}}

\newcommand{\integral}{\int\text{d}\mathbf{r}}

\graphicspath{%
    {converted_graphics/}
    {/}
    {Figures/}
}

\begin{document}



\title{Many-Body Spin Hall Effect with Space Inversion Symmetry}
\author{Nguyen Thanh Phuc}
\affiliation{Department of Theoretical and Computational Molecular Science, Institute for Molecular Science, Okazaki 444-8585, Japan}
\affiliation{Department of Structural Molecular Science, The Graduate University for Advanced Studies, Okazaki 444-8585, Japan}
\author{Masahito Ueda}
\affiliation{Department of Physics, University of Tokyo, 7-3-1 Hongo, Bunkyo-ku, Tokyo 113-0033, Japan}
\affiliation{RIKEN Center for Emergent Matter Science (CEMS), Wako, Saitama 351-0198, Japan}
\date{\today}

\begin{abstract}
In contrast to the ordinary spin Hall effect (SHE), which is a single-body phenomenon caused by the spin-orbit interaction (SOI), we propose a many-body SHE induced by the dipole-dipole interaction (DDI) between particles and demonstrate it in a system of ultracold magnetic atoms. While the SOI usually requires the breaking of space inversion symmetry, the DDI preserves it. The many-body SHE can, in principle, be observed in a wide range of systems with large dipole moments and offers a powerful tool to generate spin currents, an essential ingredient in spintronics and atomtronics.
\end{abstract}


\maketitle

\textit{Introduction.}
The spin Hall effect (SHE), first proposed for electrons by Dyakonov and Perel~\cite{Dyakonov71}, is a single-body phenomenon in which carriers undergo a spin-dependent force that deflects their trajectories in the perpendicular direction. This force is reminiscent of the Lorentz force but has the opposite directions for spin-up and spin-down particles. With the SHE, one can generate a nonzero spin current $\mathbf{j}_\mathrm{s}=\mathbf{j}_\uparrow-\mathbf{j}_\downarrow$ at zero net particle current $\mathbf{j}=\mathbf{j}_\uparrow+\mathbf{j}_\downarrow$~\cite{Sinova15}. The SHE was first observed in GaAs~\cite{Kato04, Wunderlich05, Murakami03, Sinova04} and later in a Bose-Einstein condensate (BEC) of $^{87}$Rb under a spin-dependent artificial gauge field~\cite{Beeler13} and in the propagation of light through an air-glass interface~\cite{Hosten08, Onoda04}. Here, particles with (pseudo)spin-1/2 are electrons, atoms with hyperfine spins and circularly-polarized photons. The spin current generated by the SHE has found important applications in the field of spintronics for a new generation of magnetic random access memories~\cite{Sinova17, Sinova12}. Moreover, the SHE serves as an efficient mechanism to manipulate spins in antiferromagnets, which have inherent advantages over ferromagnets as being natural materials for nonvolatile, radiation- and magnetic-field-insensitive technologies~\cite{Zelezny14, Wadley16, Olejnik17}.

The above ordinary SHE originates from the spin-orbit interaction (SOI)~\cite{Manchon15, Rashba60, Dresselhaus55}. This single-body interaction induces the SHE through either an intrinsic mechanism, in which particles experience a spin-dependent magnetic field generated directly by the SOI, or an extrinsic mechanism that requires an interplay between the SOI and impurity scattering in solid materials. The typical SOIs including the Rashba and Dresselhaus SOIs requires the breaking of space inversion symmetry (SIS). Examples include noncentrosymmetric zinc-blende or wurtzite semiconductors~\cite{Dresselhaus55, Dyakonov86} and quantum wells with broken structural inversion symmetry along the growth direction~\cite{Vasko79, Bychkov84}. In this paper, we propose a many-body SHE induced by the dipole-dipole interaction (DDI). The DDI couples the spin and orbital degrees of freedom in a way similar to the SOI~\cite{Pasquiou10, Pasquiou11, Paz13, Syzranov14} but preserves the SIS unlike the Rashba and Dresselhaus SOIs. The many-body SHE can, in principle, be observed in a diverse group of many-body systems whose constituent particles possess electric or magnetic dipole moments. Examples include ultracold heteronuclear molecules ($^{40}$K$^{87}$Rb~\cite{Ni08, Moses15, Yan13}, $^{87}$Rb$^{133}$Cs~\cite{Takekoshi14, Molony14}, $^{23}$Na$^{40}$K~\cite{Park15}, $^{23}$Na$^{87}$Rb~\cite{Guo16}), ultracold atoms with large magnetic dipole moments ($^{52}$Cr~\cite{Griesmaier05, Pasquiou10, Pasquiou11, Paz13}, $^{168}$Er~\cite{Aikawa12}, $^{164}$Dy~\cite{Lu11}), Rydberg gases~\cite{Kiffner13}, solid-state electronic materials with small or vanishing SOI, spin-triplet dimer excitations in quantum antiferromagnets~\cite{Auerbach-book}, and nuclear and electronic spins of nitrogen-vacancy (NV) centers in diamond~\cite{Jelezko04, Kucsko16}. As a concrete example, we investigate the spin dynamics of an ensemble of ultracold $^{52}$Cr atoms loaded into an optical lattice~\cite{Pasquiou11, Paz13} and show that the many-body SHE can be observed experimentally.

\textit{Many-body SHE with SIS.}
An ordinary SHE is generated by the SOI that often requires the breaking of SIS. However, the SHE itself may occur in systems with SIS. Indeed, under space inversion, both the spin current density $\mathbf{j}_\mathrm{s}$ and the applied electric field $\mathbf{E}$ change their signs, but the spin conductivity matrix $\sigma_\mathrm{s}$ remains unchanged ($\mathbf{j}_\mathrm{s}=\sigma_\mathrm{s}\mathbf{E}$). The SHE is therefore compatible with the SIS and should appear in systems where the spin and orbital degrees of freedom (DOF) are coupled to each other in a way that preserves the SIS. To this end, the interaction Hamiltonian must be even in each of these DOFs. Moreover, for a many-body SHE we need an anisotropic interaction $H(\hat{\mathbf{s}}_i,\hat{\mathbf{s}}_j,\R_{ij})$ between particles such that the spin DOFs are coupled to vectorial orbital DOFs rather than scalar ones. These requirements are satisfied by the dipole-dipole interaction (DDI)~\cite{Jackson-book}
\begin{align}
H_\mathrm{dd}(\hat{\mathbf{d}}_i, \hat{\mathbf{d}}_j, \R_{ij})=C\frac{\hat{\mathbf{d}}_i\cdot\hat{\mathbf{d}}_j-3(\hat{\mathbf{d}}_i\cdot\mathbf{e}_{ij})(\hat{\mathbf{d}}_j\cdot\mathbf{e}_{ij})}{4\pi r_{ij}^3}.
\label{eq: dipole-dipole interaction}
\end{align}
Here $r_{ij}=|\R_{ij}|$ is the distance between two dipole moments $\hat{\mathbf{d}}_i$ and $\hat{\mathbf{d}}_j$, $\mathbf{e}_{ij}=\R_{ij}/r_{ij}$, and $C=1/\epsilon_0$ ($\mu_0$) for electric (magnetic) dipoles with $\epsilon_0$ and $\mu_0$ being the vacuum permittivity and permeability, respectively. It is evident that the right-hand side of Eq.~\eqref{eq: dipole-dipole interaction} is invariant upon the space inversion transformation: $\R_i\to -\R_i$, $\R_j\to -\R_j$, and $\R_{ij}\to -\R_{ij}$. It is also clear that the DDI couples the spin and orbital degrees of freedom in such a manner that the total angular momentum is conserved. To this end, the DDI can be decomposed as $H_\mathrm{dd}(\hat{\mathbf{d}}_i, \hat{\mathbf{d}}_j, \R_{ij})=H_{ij}^{\delta L=0}+H_{ij}^{\delta L=\pm2}$~\cite{Pasquiou10, Paz13, Syzranov14}, where
\begin{align}
H_{ij}^{\delta L=0}=&\frac{C}{4\pi r_{ij}^3}\left(\frac{\hat{d}_i^{-1}\hat{d}_j^1+\hat{d}_i^{1}\hat{d}_j^{-1}}{2}+\hat{d}_i^z\hat{d}_j^z \right), \label{eq: 1st component of DD interaction}\\
H_{ij}^{\delta L=\pm2}=&-\frac{3C}{8\pi r_{ij}^3} \left(\hat{d}_i^1\hat{d}_j^1e^{-2i\phi_{ij}}+\hat{d}_i^{-1}\hat{d}_j^{-1}e^{2i\phi_{ij}}\right). \label{eq: 2nd component of DD interaction}
\end{align}
Here, $\hat{d}_j^{\pm1}\equiv \mp(\hat{d}_j^x\pm i\hat{d}_j^y)/\sqrt{2}$ are the spherical harmonic components of the dipole-moment operator $\hat{\mathbf{d}}_j$, and $(r_{ij},\phi_{ij})$ are the polar coordinates of $\R_{ij}$. While $H_{ij}^{\delta L=0}$ conserves both the spin and orbital angular momenta, $H_{ij}^{\delta L=\pm2}$ raises (lowers) the spin angular momentum by two and simultaneously lowers (raises) the orbital angular momentum by the same amount so as to conserve the total angular momentum. It is this DDI-induced spin-orbit coupling that gives rise to the  many-body SHE proposed in this paper.

\textit{Many-body SHE in ultracold magnetic atoms.}
In the following, for the sake of concreteness we consider a system of $^{52}$Cr atoms loaded into a two-dimensional (2D) optical lattice in the $xy$ plane~\cite{Pasquiou11, Paz13} (see Fig.~\ref{fig: system}). As shown below, the many-body SHE can be demonstrated in the Mott-insulator regime where atoms are strongly localized at the lattice sites, leaving only their internal degrees of freedom free. The energy degeneracy of spin states $|S=3,m_S=3,\cdots,-3\rangle$ of chromium atoms can be lifted by using a light-induced quadratic Zeeman shift (via the ac Stark effect)~\cite{Gerbier06, Leslie09}, which preserves the SIS. In the following, we restrict our consideration to the case in which the energy shift is sufficiently large so that the spin dynamics governed by the DDI essentially involves only three spin states $|m_S=1,0,-1\rangle$. In a general case that involves more spin states, the system's dynamics is more complex; however, our prediction should not change at least qualitatively if the definition of spin current is generalized to include all spin states. 

\begin{figure}[tbp] 
  \centering
  \includegraphics[width=3.3in,keepaspectratio]{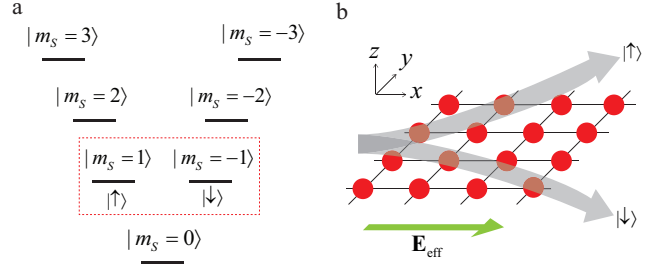}
  \caption{Many-body SHE in a system of ultracold magnetic chromium atoms. (\textbf{a}) Spin states $|m_S\rangle:=|S=3, m_S\rangle$ ($m_S=3,\cdots,-3$) of chromium atoms, where $S$ and $m_S$ denote the spin quantum number and its projection onto the quantization axis, respectively. The energy degeneracy of spin states is lifted by a light-induced quadratic Zeeman shift. (\textbf{b}) Schematic illustration of the many-body SHE. Atoms (red spheres) are placed at lattice sites of a two-dimensional optical lattice in the $xy$ plane and interact with each other by the DDI. The atomic spin dynamics is mapped to a system of moving spin-1/2 particles in which two spin states $|m_S=\pm1\rangle$ represent the spin-up and spin-down states. An effective electric field $\mathbf{E}_\mathrm{eff}$ (green arrow) drives the particles in the $x$ direction, and spin-up and spin-down particles are deflected oppositely towards the $+y$ and $-y$ directions, respectively. This many-body SHE produces a nonzero spin current at zero net particle current, which can directly be measured with a Stern-Gerlach experiment.}
  \label{fig: system}
\end{figure}

If the fraction of atoms in the excited states $|m_S=\pm1\rangle$ is small, the last term ($\hat{d}_i^z\hat{d}_j^z$) in Eq.~\eqref{eq: 1st component of DD interaction} becomes negligibly small compared with the other terms in the DDI. In this case, the spin dynamics of the atomic ensemble can be mapped onto the dynamics of a system of free spin-1/2 particles where the two spin states $|m_S=1\rangle$ and $|m_S=-1\rangle$ correspond to the spin-up and spin-down states, respectively, while the ground state $|m_S=0\rangle$ represents the vacuum for those particles. From the energy conservation, it is clear that the term $\hat{d}^1_i\hat{d}^1_j$ in Eq.~\eqref{eq: 2nd component of DD interaction} makes the state transition at site $i$ ($j$) from $|m_S=-1\rangle$ ($|m_S=0\rangle$) to $|m_S=0\rangle$ ($|m_S=1\rangle$), which implies that the state $|m_S=-1\rangle$ is transferred from $i$ to $j$ followed by a spin flip $|m_S=-1\rangle \to |m_S=1\rangle$. Similarly, the term $\hat{d}^{-1}_i\hat{d}^{-1}_j$ causes the state transition at site $i$ ($j$) from $|m_S=1\rangle$ ($|m_S=0\rangle$) to $|m_S=0\rangle$ ($|m_S=-1\rangle$), which implies that the state $|m_S=1\rangle$ is transferred from $i$ to $j$ followed by a spin flip $|m_S=1\rangle \to |m_S=-1\rangle$. The second-quantized DDI Hamiltonian of the system then reads~\cite{Syzranov14}
\begin{align}
H_\mathrm{dd}=t_0\sum_{\substack{i,j\not=i\\ \sigma=\uparrow\downarrow}}\frac{\hat{b}^\dagger_{i\sigma}\hat{b}_{j\sigma}}{r_{ij}^3}+t_2\sum_{i,j\not=i}\frac{e^{-2i\phi_{ij}}\hat{b}^\dagger_{i\uparrow}\hat{b}_{j\downarrow}+e^{2i\phi_{ij}}\hat{b}^\dagger_{i\downarrow}\hat{b}_{j\uparrow}}{r_{ij}^3},
\label{eq: hopping picture}
\end{align}
where $\hat{b}_{j\sigma}$ ($\hat{b}^\dagger_{j\sigma}$) is the annihilation (creation) operator of a particle with spin $\sigma=\uparrow,\downarrow$ at lattice site $j$. The first term on the right-hand side of Eq.~\eqref{eq: hopping picture} describes a spin-conserving tunneling, while the second term describes tunneling accompanied by a spin flip, where $t_0=-3Cd^2/(4\pi)$, $t_2=3t_0$, and $d=g\mu_\mathrm{B}$ with $g$ and $\mu_\mathrm{B}$ being the Land\'{e} $g$-factor and the Bohr magneton, respectively. It is noteworthy that the effective Hamiltonian~\eqref{eq: hopping picture}, which is obtained by neglecting the last term in Eq.~\eqref{eq: 1st component of DD interaction}, also preserves the SIS as the original Hamiltonian~\eqref{eq: dipole-dipole interaction}. In fact, the space inversion transformation makes $\phi_{ij}\to \phi_{ij}+\pi$ and in turn $e^{\pm 2i\phi_{ij}}$ invariant. Due to the translation invariance of the system, $H_\mathrm{dd}$ can be expressed as $H_\mathrm{dd}=\sum_\K \begin{pmatrix} \hat{b}_{\K\uparrow}^\dagger, \hat{b}_{\K\downarrow}^\dagger \end{pmatrix} H \begin{pmatrix} \hat{b}_{\K\uparrow}\\ \hat{b}_{\K\downarrow} \end{pmatrix}$, where
\begin{align}
H=\begin{pmatrix}
t_0f^{(0)}(\K) & t_2f^{(2)}(\K) \\
t_2f^{(2)}(\K)^* & t_0f^{(0)}(\K) 
\end{pmatrix},
\label{eq: Hamiltonian matrix element}
\end{align}
$\hat{b}_{\K\sigma}\equiv (1/N^2)\sum_j e^{-i\K\cdot\R_j}\hat{b}_{j\sigma}$ ($N$ is the number of lattice sites in one direction), $f^{(0)}(\K)=N^2\sum_{\R_j\not=0}e^{-i\K\cdot\R_j}/r_j^3$ and $f^{(2)}(\K)=N^2\sum_{\R_j\not=0}e^{-i\K\cdot\R_j}e^{2i\phi_j}/r_j^3$. The diagonalization of Eq.~\eqref{eq: Hamiltonian matrix element} gives
\begin{align}
H=E(\mathbf{0})\left(\hat{b}^\dagger_{\mathbf{0}\uparrow}\hat{b}_{\mathbf{0}\uparrow}+\hat{b}^\dagger_{\mathbf{0}\downarrow}\hat{b}_{\mathbf{0}\downarrow}\right)+\sum_{\substack{\K\not=\mathbf{0}\\ \xi=\pm}} E_{\xi}(\K) \hat{b}^\dagger_{\K\xi}\hat{b}_{\K\xi},
\label{eq: diagonalized Hamiltonian}
\end{align}
where $E(\mathbf{0})=t_0f^{(0)}(\K=\mathbf{0})$, $E_\pm(\K)=t_0f^{(0)}(\K)\pm t_2|f^{(2)}(\K)|$ and $\hat{b}_{\K+}=(e^{i\theta_\K}\hat{b}_{\K\uparrow}+\hat{b}_{\K\downarrow})/\sqrt{2}, \hat{b}_{\K-}=(\hat{b}_{\K\uparrow}-e^{-i\theta_\K}\hat{b}_{\K\downarrow})/\sqrt{2}$ with $\theta_\K$ being the phase of $f^{(2)}(\K)$. In the following, we show that the SHE emerges if an external force is applied to these spin-1/2 particles.

\textit{Spin Hall conductivity.}
Since the spin-up and spin-down particles correspond to the $|m_S=\pm1\rangle$ spin states of the atoms, a force exerted on those particles can be generated by, for example, a spatial gradient of the light intensity that induces a quadratic Zeeman shift in the energy levels. This force plays the role of an effective electric field for the spin-1/2 particles, which can be expressed in terms of an effective vector potential as $\mathbf{E}_\mathrm{eff}(\R,t)=-\partial \mathbf{A}_\mathrm{eff}(\R,t)/\partial t$. In the presence of a vector potential, the Hamiltonian $H$ of the system is obtained through the replacement of the momentum operator $\hat{\mathbf{p}}$ with $\hat{\mathbf{p}}-\mathbf{A}_\mathrm{eff}(\hat{\R},t)$ to ensure the local gauge invariance. The first-order perturbation of the Hamiltonian is given by $\delta H=-\int \text{d}\R \;\hat{\mathbf{j}}^\mathrm{p}(\R)\cdot\mathbf{A}_\mathrm{eff}(\R,t)$, where $\hat{\mathbf{j}}^\mathrm{p}(\R)=\frac{1}{2}\left\{\hat{n}(\R),\hat{\mathbf{v}}\right\}$ is the paramagnetic component of the particle-current density operator~\cite{Rammer-book}. Here, the curly brackets denote the anticommutator and the particle-density and group-velocity operators are given by $\hat{n}(\R)=\delta(\R-\hat{\R})$ and $\hat{\mathbf{v}}=\frac{\partial H}{\partial \hat{\p}}$, respectively. From Eq.~\eqref{eq: Hamiltonian matrix element}, the group-velocity operator is found to be
\begin{align}
\hat{\mathbf{v}}=\frac{1}{\hbar}\begin{pmatrix}
t_0\mathbf{g}^{(0)}(\hat{\mathbf{p}}/\hbar) & t_2\mathbf{g}^{(2)}(\hat{\mathbf{p}}/\hbar) \\
t_2\mathbf{g}^{(2)}(\hat{\mathbf{p}}/\hbar)^* & t_0\mathbf{g}^{(0)}(\hat{\mathbf{p}}/\hbar) 
\end{pmatrix},
\end{align}
where $\mathbf{g}^{(0),(2)}(\K)\equiv\nabla_\K f^{(0),(2)}(\K)$. The total particle-current density is given by the sum of the paramagnetic and diamagnetic components, where the latter is given by the product of the particle density and the vector potential. In the linear-response regime, the paramagnetic particle-current density in the frequency domain is given by $j^\mathrm{p}_\alpha(\R,\omega)=\sum_\beta \integral'\,K_{\alpha,\beta}(\R,\R',\omega)A_\beta(\R',\omega)$, where $\alpha,\beta=x,y,z$ and $K_{\alpha,\beta}(\R,\R',\omega)$ is the paramagnetic response function. Since the total particle current should vanish at the zero-frequency limit, where the vector potential becomes time independent, leading to the vanishing electric field, the diamagnetic current can be expressed in terms of the paramagnetic response function $K_{\alpha,\beta}(\R,\R',\omega=0)$ at zero frequency. Summing the two components, we obtain the total particle-current density as $j_\alpha(\R,\omega)=\sum_\beta \integral'\,\left[K_{\alpha,\beta}(\R,\R',\omega)-K_{\alpha,\beta}(\R,\R',\omega=0)\right]A_\beta(\R',\omega)$. Now, the spin-current density operator with the $z$-axis polarization is given by $\hat{\mathbf{j}}_\mathrm{s}^z(\R)= \frac{\hbar}{4}\left\{\hat{\sigma}_z,\hat{\mathbf{j}}(\R)\right\}$, where $\hat{\sigma}_z=\text{diag}(1,-1)$ is the Pauli matrix~\cite{Sinova04}. This spin current can be measured directly by the Stern-Gerlach experiment~\cite{Gerlach22}. As the vector potential is related to the electric field in the frequency domain by $\mathbf{A}(\R,\omega)=\mathbf{E}(\R,\omega)/(i\omega)$, the spin conductivity matrix is given by $(\chi_\mathrm{s})_{\alpha\beta}(\R,\R',\omega)=[(K_\mathrm{s})_{\alpha,\beta}(\R,\R',\omega)-(K_\mathrm{s})_{\alpha,\beta}(\R,\R',\omega=0)]/(i\omega)$,
where $(K_\mathrm{s})_{\alpha,\beta}(\R,\R',\omega)$ is the paramagnetic response function of the spin current.

\begin{figure}[tbp] 
  \centering
  \includegraphics[width=2in,keepaspectratio]{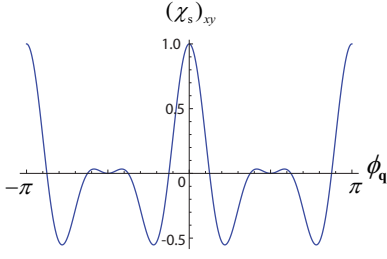}
  \caption{Dependence of the spin Hall conductivity $(\chi_\mathrm{s})_{xy}$ on the direction of the initial excitation with wavevector $\mathbf{q}$. Here $\mathbf{q}=(q,\phi_\mathbf{q})$ are the polar coordinates, and $(\chi_\mathrm{s})_{xy}$ is shown in units of $1/[\pi(Nqa)^2]$, where $a$ and $N$ are the lattice constant and the number of lattice sites in a linear dimension, respectively.}
  \label{fig: angular dependence}
\end{figure}

As an initial state that preserves the SIS, we consider the case in which half of the spin-1/2 particles are initially prepared in the eigenstate $|\Psi_1\rangle=|\mathbf{q},+\rangle$ of the Hamiltonian~\eqref{eq: diagonalized Hamiltonian} with a momentum $\hbar\mathbf{q}\not=0$ and the other half in the eigenstate $|\Psi_2\rangle=|-\mathbf{q},+\rangle$. Such an initial state can be realized, for example, by using a pair of independent counter-propagating laser beams with appropriately chosen directions and polarizations to excite the atomic ensemble. Note that in contrast to the case of the Rashba SOI, here the spin components of $|\mathbf{q},+\rangle$ and $|-\mathbf{q},+\rangle$ are identical, reflecting the fact that the system preserves the SIS. Using the linear-response theory, the paramagnetic response function is found to be
\begin{widetext}
\begin{align}
(K_\mathrm{s})_{\alpha\beta}(\R,\R',\omega)=&-\frac{1}{4\pi}\sum_{j=1}^2\sum_n \left[\frac{\langle \Psi_j|(\hat{j}_\mathrm{s}^\mathrm{p})^z_\alpha(\R)|\psi_n\rangle\langle \psi_n|\hat{j}^\mathrm{p}_\beta(\R')|\Psi_j\rangle}{E_0-E_n-\hbar\omega+i\eta}
+\frac{\langle \psi_n|(\hat{j}_\mathrm{s}^\mathrm{p})^z_\alpha(\R)|\Psi_j\rangle\langle \Psi_j|\hat{j}^\mathrm{p}_\beta(\R')|\psi_n\rangle}{E_0-E_n+\hbar\omega-i\eta}\right],
\label{eq: paramagnetic response function}
\end{align}
\end{widetext}
where $\eta$ is an infinitesimal positive number and the sum is taken over all eigenstates $|\psi_n\rangle$ of Hamiltonian~\eqref{eq: diagonalized Hamiltonian}. Here $E_0$ is the eigenenergy of $|\Psi_{1,2}\rangle$ and $E_n$ is that of $|\psi_n\rangle$. The matrix elements in Eq.~\eqref{eq: paramagnetic response function} can be evaluated straightforwardly. As the unperturbed system is translation invariant, both $(K_\mathrm{s})_{\alpha\beta}(\R,\R',\omega)$ and $(\chi_\mathrm{s})_{\alpha\beta}(\R,\R',\omega)$ are functions of $\R-\R'$. By making the Fourier transformation $(\chi_\mathrm{s})_{\alpha\beta}(\K,\omega)=\int \text{d}(\R-\R') e^{-i\K\cdot(\R-\R')}(\chi_\mathrm{s})_{\alpha\beta}(\R-\R',\omega)$ and taking the limits of $\K\to \mathbf{0}$ and $\omega\to 0$, which corresponds to a homogeneous and static driving force, we obtain
\begin{widetext}
\begin{align}
(\chi_\mathrm{s})_{\alpha\beta}(\K\to \mathbf{0},\omega\to 0)=&\frac{t_0t_2}{2\pi S}\sum\limits_{\p=\pm\mathbf{q}}
\frac{g^{(0)}_\alpha(\p)\mathrm{Im}\Big\{e^{i\theta_{\p}}\left[-g^{(2)}_\beta(\p)+e^{2i\theta_{\p}}g^{(2)}_\beta(\p)^*\right]\Big\}}{[E_+(\p)-E_-(\p)]^2},
\label{eq: homogeneous, static limit}
\end{align}
\end{widetext}
where $S=(Na)^2$ is the area of the lattice. In the long-wavelength limit $ka\ll 1$ ($a$ is the lattice constant) and for the square lattice, $f^{(0)}(\K)$ and $f^{(2)}(\K)$ have the asymptotic forms of $a^3f^{(0)}(\K)/N^2\simeq A-2\pi ka$ and $a^3f^{(2)}(\K)/N^2\simeq 2\pi i ka e^{2i\phi_\K}/3$, where $A\simeq 9.03$ is a constant and $\phi_\K$ is the azimuthal angle of $\K$. Therefore, for $qa\ll 1$ the spin Hall conductivity $(\chi_\mathrm{s})_{xy}(\K\to\mathbf{0},\omega\to 0)$ reduces to
\begin{align}
(\chi_\mathrm{s})_{xy}=\frac{\cos(4\phi_\mathbf{q})\cos^2(\phi_\mathbf{q})}{\pi (Nqa)^2}.
\label{eq: analytic spin Hall conductivity}
\end{align}
The dependence of $(\chi_\mathrm{s})_{xy}$ on the direction of the initial momentum (angle $\phi_\mathbf{q}$) is shown in Fig.~\ref{fig: angular dependence}, demonstrating the characteristic anisotropy of the DDI [Eq.~\eqref{eq: dipole-dipole interaction}] and the $\pi$-periodicity which arises from the fact that two quanta of the angular momentum are transferred between the spin and orbital degrees of freedom by the DDI.

\textit{Numerical simulation.}
In deriving the spin Hall conductivity analytically [Eq.~\eqref{eq: analytic spin Hall conductivity}], we have assumed that the fraction of atomic excitations, i.e., the ratio of the number of atoms in the $|m_S=\pm1\rangle$ spin states to that of atoms in the $|m_S=0\rangle$ ground state, is sufficiently small so that the interaction between excitations [the last term in Eq.~\eqref{eq: 1st component of DD interaction}] can be neglected. As numerically demonstrated below, this assumption can be removed without changing qualitatively the physics we are interested in.

\begin{figure}[tbp] 
  \centering
  \includegraphics[width=3in,keepaspectratio]{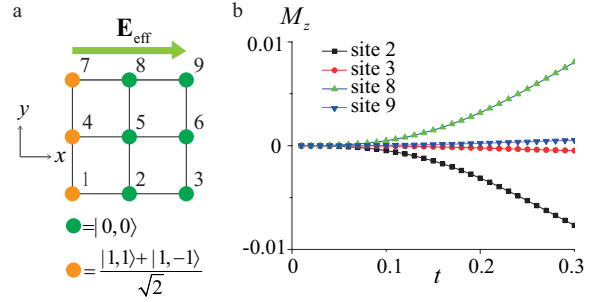}
  \caption{Dynamical simulation of the many-body SHE in a system of ultracold magnetic atoms. (\textbf{a}) Initially, all atoms are prepared in the $|m_S=0\rangle$ ground state (green circles). Atoms at the left edge are then excited to the superposition state: $(|m_S=1\rangle+|m_S=-1\rangle) /\sqrt{2}$ (orange circles). An effective electric field $\mathbf{E}_\mathrm{eff}$ is applied in the $x$ direction, whose magnitude is taken to be $E_\mathrm{eff}a=\epsilon_\mathrm{dd}/4$, where $\epsilon_\mathrm{dd}=Cd^2/(4\pi a^3)$ is the characteristic energy scale of the DDI. (\textbf{b}) Time evolutions of the magnetization $M_z(\R_j)=\langle \hat{P}^1_j-\hat{P}^{-1}_j\rangle$ of molecules at four lattice sites. Here, $\hat{P}^{\pm1}_j$ are the projection operators onto the spin states $|m_S=\pm1\rangle$ at lattice site $j$. The time $t$ is measured in units of $\hbar/\epsilon_\mathrm{dd}$.}
  \label{fig: time evolution}
\end{figure}

In the numerical simulation, atoms are located at the sites of a $3\times 3$ square lattice with unit filling, i.e., one atom per site. Initially, all atoms are prepared in the $|m_S=0\rangle$ ground state. Atoms at the left edge are then excited to the superposition state: $(|m_S=1\rangle+|m_S=-1\rangle) /\sqrt{2}$ (see Fig.~\ref{fig: time evolution}a). An effective electric field $\mathbf{E}_\mathrm{eff}$ for those excitations is applied in the $x$ direction as described by an energy potential $H_E=E_\mathrm{eff}\sum_j x_j \hat{P}^0_j$, where $x_j$ and $\hat{P}^0_j$ are the $x$-coordinate and the projection operator onto the $|m_S=0\rangle$ ground state of the atom at lattice site $j$, respectively. The time evolution of the system is obtained by using the exact diagonalization method.
 
Figure~\ref{fig: time evolution}b shows the short-time evolutions of the magnetizations $M_z(x_j,y_j)=\langle \hat{P}^1_j-\hat{P}^{-1}_j\rangle$ of atoms at four lattice sites $(x_j,y_j)=(1,2), (1,3), (3,2)$ and (3,3). Here, $\hat{P}^{\pm1}_j$ are the projection operators onto the spin states $|m_S=\pm1\rangle$ of an atom at lattice site $j$. It is evident that $M_z$ negatively increases for atoms at the bottom edge and positively increases for atoms at the top edge, while their absolute values are almost equal. These spin accumulations offer the evidence of the SHE and can directly be measured experimentally. The time scale of the spin Hall dynamics shown in Fig.~\ref{fig: time evolution}b is given by $\tau=\hbar/\epsilon_\mathrm{dd}$, where $\epsilon_\mathrm{dd}=Cd^2/(4\pi a^3)$ is the characteristic energy scale of the DDI. Using the parameters of the $^{52}$Cr~\cite{Lahaye09}, we find that $\tau\simeq 25$ ms, which makes the many-body SHE observable in current experiments. 

\textit{Concluding remarks.}
We have proposed the many-body SHE induced by the DDI between particles, which is in marked contrast to the ordinary SHE that is a single-body phenomenon generated by the SOI. We have demonstrated both analytically and numerically the many-body SHE in a system of ultracold magnetic chromium atoms. Moreover, the observation of many-body SHE, in general, does not require a Bose-Einstein condensate~\cite{Oshima16}.

Although the spin and orbital degrees of freedom are coupled to each other by both the DDI and SOI, the couplings themselves are totally different, leading to important differences between the SHEs caused by the two interactions. While the SOI is purely a single-body effect, the coupling induced by the DDI is of an intrinsic many-body nature. Whereas the ordinary SHE with SOI often requires SIS breaking, the many-body SHE induced by the DDI can occur in systems with SIS. Moreover, in contrast to the isotropic SOI, the characteristic anisotropy of the DDI is clearly reflected in the strong dependence of the spin Hall conductivity on the direction of the initial momentum. It is also noteworthy that the many-body SHE is fundamentally different from the magnon Hall effect~\cite{Katsura10, Onose10, Matsumoto11} which typically arises from a nonzero Dzyaloshinskii-Moriya interaction that breaks the SIS and plays a role similar to the SOI in the ordinary SHE. The magnon Hall effect can also be observed in the case of magnetostatic spin waves~\cite{Matsumoto11b} where the restriction of the action of the magnetic DDI between spins to one spatial dimension results in the demagnetizing field that plays a role similar to the single-body SOI in the ordinary SHE. In contrast, there is no restriction on the action of the DDI that constitutes the many-body SHE which works in general even in the strongly correlated regime where the concept of quasiparticles is no longer valid.

In the case of clean systems such as ultracold atoms and molecules, the divergence of the longitudinal conductivity makes it difficult to reach a steady state. The spin Hall conductivity, however, can be obtained by extrapolating from the time evolution of magnetization at the edges of the system (see Fig.~\ref{fig: time evolution}). In contrast, the steady state should be reached in solid-state electronic systems due to the impurity scattering, which makes the spin Hall conductivity directly measurable. In considering the effect of impurity scattering, however, it is necessary to take account of the vertex correction, which can dramatically change the spin Hall conductivity as shown for the ordinary SHE with the Rashba SOI~\cite{Inoue04}. In this case, a similar definition of the spin Hall angle~\cite{Sinova15} can be introduced to characterize the dimensionless magnitude of the many-body SHE.

Concerning the strength of interaction, although the magnitude of magnetic DDI is generally smaller than that of the electric DDI by a factor of $\alpha^2$, where $\alpha=e^2/(4\pi\epsilon_0\hbar c)$ is the fine structure constant, the magnetic DDI and the electric SOI share the same scaling with respect to the fine structure constant. Therefore, the proposed many-body SHE can find wide application not only in systems whose particles possess an electric dipole moment but also in numerous space-inversion-symmetric systems with magnetic dipole moments where the SOI is negligibly small. 

Using the SOI, the single-body quantum SHE characterized by a unidirectional edge spin transport, i.e., edge states with opposite spins propagating in opposite directions, has been observed in both electronic~\cite{Kane05, Bernevig06} and photonic~\cite{Bliokh15} materials, and it gives rise to a new and important class of materials, namely topological insulators~\cite{Hasan10, Qi11}. Similarly, the many-body SHE can open new avenues for studies of as yet unexplored classes of many-body topological materials in which inter-particle interactions play an essential role.  

\begin{acknowledgements}
N. T. Phuc thanks N. Tsuji, T. Momoi, and S. Choi for insightful discussions. This work was supported by KAKENHI Grant No. JP18H01145 from the Japan Society for the Promotion of Science, and a Grant-in-Aid for Scientific Research on Innovation Areas \textquotedblleft Topological Materials Science" (KAKENHI Grant No. JP15H05855), and the Photon Frontier Network Program from MEXT of Japan. 
\end{acknowledgements}



\begin{thebibliography}{100}
\bibitem{Dyakonov71} M. I. Dyakonov, V. I. Perel, Possibility of orienting electron spins with current. Sov. Phys. JETP \textbf{13}, 467--469 (1971).
\bibitem{Sinova15} J. Sinova, S. O. Valenzuela, J. Wunderlich, C. H. Back, T. Jungwirth, Spin Hall effects. Rev.
Mod. Phys. \textbf{87}, 1213 (2015).
\bibitem{Kato04} Y. K. Kato, R. C. Myers, A. C. Gossard, D. D. Awschalom, Observation of the spin Hall effect in semiconductors. Science \textbf{306}, 1910--1913 (2004).
\bibitem{Wunderlich05} J. Wunderlich, B. Kaestner, J. Sinova, T. Jungwirth, Experimental observation of the spin-Hall effect in a two-dimensional spin-orbit coupled semiconductor system. Phys. Rev. Lett. \textbf{94}, 047204 (2005).
\bibitem{Murakami03} S. Murakami, N. Nagaosa, S. C. Zhang, Dissipationless quantum spin current at room temperature. Science \textbf{301}, 1348 (2003).
\bibitem{Sinova04} J. Sinova, D. Culcer, Q. Niu, N. A. Sinitsyn, T. Jungwirth, A. H. MacDonald, Universal Intrinsic Spin Hall Effect. Phys. Rev. Lett. \textbf{92}, 126603 (2004).
\bibitem{Beeler13} M. C. Beeler, R. A. Williams, K. Jimenez-Garcia, L. J. LeBlanc, A. R. Perry, I. B. Spielman, The spin Hall effect in a quantum gas. Nature \textbf{498}, 201 (2013).
\bibitem{Hosten08} O. Hosten, P. Kwiat, Observation of the spin Hall effect of light via weak measurements. Science \textbf{319}, 787--790 (2008).
\bibitem{Onoda04} M. Onoda, S. Murakami, N. Nagaosa, Hall Effect of Light. Phys. Rev. Lett. \textbf{93}, 083901 (2004).
\bibitem{Sinova17} J. Sinova, T. Jungwirth, Surprises from the spin Hall effect. Physics Today \textbf{70}, 7, 38 (2017).
\bibitem{Sinova12} J. Sinova, I. Zutic, New moves of the spintronics tango. Nat. Mater. \textbf{11}, 368 (2012).
\bibitem{Zelezny14} J. Zelezny, H. Gao, K. Vyborny, J. Zemen, J. Masek, A. Manchon, A. Wunderlich, J. Sinova, T. Jungwirth, Relativistic Neel-Order Fields Induced by Electrical Current in Antiferromagnets
Phys. Rev. Lett. \textbf{113}, 157201 (2014).
\bibitem{Wadley16} P. Wadley, B. Howells, J. Zelezny, C. Andrews, V. Hills, R. P. Campion, V. Novak, K. Olejnik, F. Maccherozzi, S. S. Dhesi, S. Y. Martin, T. Wagner, J. Wunderlich, F. Freimuth, Y. Mokrousov, J. Kunes, J. S. Chauhan, M. J. Grzybowski, A. W. Rushforth, K. W. Edmonds, B. L. Gallagher, T. Jungwirth, Electrical switching of an antiferromagnet. Science \textbf{351}, 587 (2016).
\bibitem{Olejnik17} K. Olejnik, V. Schuler, X. Marti, V. Novak, Z. Kaspar, P. Wadley, R. P. Campion, K. W. Edmonds, B. L. Gallagher, J. Garces, M. Baumgartner, P. Gambardella, T. Jungwirth, Antiferromagnetic CuMnAs multi-level memory cell with microlectronic compatibility. Nat. Comm. \textbf{8}, 15434 (2017).
\bibitem{Manchon15} A. Manchon, H. C. Koo, J. Nitta, S. M. Frolov, R. A. Duine, New perspectives for Rashba spin–orbit coupling, Nat. Mat. \textbf{14}, 871 (2015).
\bibitem{Rashba60} E. Rashba, Properties of semiconductors with an extremum loop. 1. Cyclotron and combinational resonance in a magnetic field perpendicular to the plane of the loop. Sov. Phys. Solid State \textbf{2}, 1109--1122 (1960).
\bibitem{Dresselhaus55} G. Dresselhaus, Spin-orbit coupling effects in zinc blende structures, Phys. Rev. \textbf{100}, 580--586 (1955).
\bibitem{Dyakonov86} M. I. Dyakonov, V. Y. Kachorovskii, Spin relaxation of two-dimensional electrons in noncentrosymmetric semiconductors. Sov. Phys. Semicond. \textbf{20}, 110–112 (1986).
\bibitem{Vasko79} F. T. Vasko, Spin splitting in the spectrum of two-dimensional electrons due to the surface potential. P. Zh. Eksp. Teor. Fiz. \textbf{30}, 574–577 (1979).
\bibitem{Bychkov84} Y. A. Bychkov, E. I. Rasbha, Properties of a 2D electron gas with lifted spectral degeneracy. P. Zh. Eksp. Teor. Fiz. \textbf{39}, 66–69 (1984).
\bibitem{Pasquiou10} B. Pasquiou, G. Bismut, Q. Beaufils, A. Crubellier, E. MarLechal, P. Pedri, L. Vernac, O. Gorceix, B. Laburthe-Tolra, Control of dipolar relaxation in external fields, Phys. Rev. A \textbf{81}, 042716 (2010).
\bibitem{Pasquiou11} B. Pasquiou, G. Bismut, E. Marechal, P. Pedri, L. Vernac, O. Gorceix, and B. Laburthe-Tolra, Spin Relaxation and Band Excitation of a Dipolar Bose-Einstein Condensate in 2D Optical Lattices, Phys. Rev. Lett. \textbf{106}, 015301 (2011).
\bibitem{Paz13} A. de Paz, A. Chotia, E. MarLechal, P. Pedri, L. Vernac, O. Gorceix, B. Laburthe-Tolra, Resonant demagnetization of a dipolar Bose-Einstein condensate in a three-dimensional optical lattice, Phys. Rev. A \textbf{87}, 051609 (2013).
\bibitem{Syzranov14} S. V. Syzranov, M. L. Wall, V. Gurarie, A. M. Rey, Spin-orbital dynamics in a system of polar molecules, Nat. Comm. \textbf{5}, 5391 (2014).
\bibitem{Ni08} K.-K. Ni, S. Ospelkaus, M. H. G. de Miranda, A. Pefer, B. Neyenhuis, J. J. Zirbel, S. Kotochigova, P. S. Julienne, D. S. Jin, J. Ye, A High Phase-Space-Density Gas of Polar Molecules, Science \textbf{322}, 231 (2008).
\bibitem{Moses15} S. A. Moses, J. P. Covey, M. T. Miecnikowski, B. Yan, B. Gadway, J. Ye, D. S. Jin, Creation of a low-entropy quantum gas of polar molecules in an optical lattice, Science \textbf{350}, 659 (2015).
\bibitem{Yan13} B. Yan, S. A. Moses, B. Gadway, J. P. Covey, K. R. A. Hazzard, A. M. Rey, D. S. Jin, J. Ye, Observation of dipolar spin-exchange interactions with lattice-confined polar molecules, Nature \textbf{501}, 521 (2013).
\bibitem{Takekoshi14} T. Takekoshi, L. Reichsollner, A. Schindewolf, J. M. Hutson, C. R. L. Sueur,
O. Dulieu, F. Ferlaino, R. Grimm, H.-C. Nagerl, Ultracold dense samples of dipolar RbCs molecules in the rovibrational and hyperfine ground state. Phys. Rev. Lett. \textbf{113}, 205301 (2014).
\bibitem{Molony14} P. K. Molony, P. D. Gregory, Z. Ji, B. Lu, M. P. Koppinger, C. R. L. Sueur,
C. L. Blackley, J. M. Hutson, S. L. Cornish, Creation of ultracold $^{87}$Rb$^{133}$Cs molecules in the rovibrational ground state. Phys. Rev. Lett. \textbf{113}, 255301 (2014).
\bibitem{Park15} J. W. Park, S. A. Will, M. W. Zwierlein, Ultracold dipolar gas of fermionic $^{23}$Na$^{40}$K molecules in their absolute ground state. Phys. Rev. Lett. \textbf{114}, 205302 (2015).
\bibitem{Guo16} M. Guo, B. Zhu, B. Lu, X. Ye, F. Wang, R. Vexiau, N. Bouloufa-Maafa,
G. Quemener, O. Dulieu, D. Wang, Creation of an ultracold gas of ground-state dipolar $^{23}$Na$^{87}$Rb molecules. Phys. Rev. Lett. \textbf{116}, 205303 (2016).
\bibitem{Griesmaier05} A. Griesmaier, J. Werner, S. Hensler, J. Stuhler, T. Pfau, Bose-Einstein Condensation of Chromium, Phys. Rev. Lett. \textbf{94}, 160401 (2005). 
\bibitem{Aikawa12} K. Aikawa, A. Frisch, M. Mark, S. Baier, A. Rietzler, R. Grimm, F. Ferlaino, Bose-Einstein Condensation of Erbium, Phys. Rev. Lett. \textbf{108}, 210401 (2012).
\bibitem{Lu11} M. Lu, N. Q. Burdick, S. H. Youn, B. L. Lev, Strongly Dipolar Bose-Einstein Condensate of Dysprosium, Phys. Rev. Lett. \textbf{107}, 190401 (2011).
\bibitem{Kiffner13} M. Kiffner, W. Li, D. Jaksch, Three-Body Bound States in Dipole-Dipole Interacting Rydberg Atoms, Phys. Rev. Lett. \textbf{111}, 233003 (2013).
\bibitem{Auerbach-book} S. Auerbach, \textit{Interacting Electrons and Quantum Magnetism} (Springer, New York, 1994).
\bibitem{Jelezko04} F. Jelezko, T. Gaebel, I. Popa, M. Domhan, A. Gruber, J. Wrachtrup, Observation of Coherent Oscillation of a Single Nuclear Spin and Realization of a Two-Qubit Conditional Quantum Gate, Phys. Rev. Lett. \textbf{93}, 130501 (2004).
\bibitem{Kucsko16} G. Kucsko, S. Choi, J. Choi, P. C. Maurer, H. Sumiya, S. Onoda, J. Isoya, F. Jelezko, E. Demler, N. Y. Yao, M. D. Lukin, Critical thermalization of a disordered dipolar spin system in diamond, arxiv:1609.08216 (2016).
\bibitem{Jackson-book} J. D. Jackson, \textit{Classical Electrodynamics} (John Wiley \& Sons, ed. 3, 1999). 
\bibitem{Gerbier06} F. Gerbier, A. Widera, S. Folling, O. Mandel, and I. Bloch, Resonant control of spin dynamics in ultracold quantum gases by microwave dressing, Phys. Rev. A \textbf{73}, 041602 (2006).
\bibitem{Leslie09} S. R. Leslie, J. Guzman, M. Vengalattore, J. D. Sau, M. L. Cohen, and D. M. Stamper-Kurn, Amplification of fluctuations in a spinor Bose-Einstein condensate, Phys. Rev. A \textbf{79}, 043631 (2009).

\bibitem{Rammer-book} J. Rammer, \textit{Quantum Transport Theory} (Perseus Books, 1998).
\bibitem{Gerlach22} W. Gerlach, O. Stern, Der experimentelle Nachweis der Richtungsquantelung im Magnetfeld. Zeitschrift für Physik \textbf{9}, 349–352 (1922).
\bibitem{Lahaye09} T. Lahaye, C. Menotti, L. Santos, M. Lewenstein, and T. Pfau, The physics of dipolar bosonic quantum
gases, Rep. Prog. Phys. \textbf{72}, 126401 (2009).

\bibitem{Oshima16} T. Oshima, Y. Kawaguchi, Spin Hall effect in a spinor dipolar Bose-Einstein condensate. Phys. Rev. A \textbf{93}, 053605 (2016).
\bibitem{Katsura10} H. Katsura, N. Nagaosa, and P. A. Lee, Theory of the thermal Hall effect in quantum magnets. Phys. Rev. Lett. \textbf{104}, 066403 (2010).
\bibitem{Onose10} Y. Onose et al., Observation of the magnon Hall effect. Science \textbf{329}, 297 (2010).
\bibitem{Matsumoto11} R. Matsumoto and S. Murakami, Theoretical prediction of a rotating magnon wave packet in ferromagnets. Phys. Rev. Lett. \textbf{106}, 197202 (2011).
\bibitem{Matsumoto11b} R. Matsumoto and S. Murakami, Rotational motion of magnons and the thermal Hall effect. Phys. Rev. B \textbf{84}, 184406 (2011).
\bibitem{Inoue04} J. Inoue, G. E. Bauer, and L. W. Molenkamp, Suppression of the persistent spin Hall current by defect scattering, Phys. Rev. B \textbf{70}, 041303 (2004).
\bibitem{Kane05} C. L. Kane, E. J. Mele, $Z_2$ Topological Order and the Quantum Spin Hall Effect. Phys. Rev. Lett. \textbf{95}, 146802 (2005).
\bibitem{Bernevig06} B. A. Bernevig, T. L. Hughes, S. C. Zhang, Quantum spin Hall effect and topological phase transition in HgTe quantum wells. Science \textbf{314}, 1757--1761 (2006).
\bibitem{Bliokh15} K. Y. Bliokh, D. Smirnova, F. Nori, Quantum spin Hall effect of light. Science \textbf{348}, 1448 (2015).
\bibitem{Hasan10} M. Z. Hasan, C. L. Kane, Colloquium: Topological insulators. Rev. Mod. Phys. \textbf{82}, 3045--3067 (2010).
\bibitem{Qi11} X.-L. Qi, S.-C. Zhang, Topological insulators and superconductors. Rev. Mod. Phys. \textbf{83}, 1057--1110 (2011).
\end{thebibliography}
\end{document}